\documentclass[aps,preprint,groupedaddress,nofootinbib]{revtex4-1}
 
\usepackage{graphicx}
\usepackage[caption=false]{subfig}
\usepackage{multirow}
\usepackage{array}
\usepackage{adjustbox}
 \usepackage{amsmath}
\usepackage{lineno, blindtext, color}
 \usepackage{float}
  \usepackage{tikz}
  \usepackage{slashed}
\newcommand{\as}{\alpha_s}
\newcommand{\vcb}{|V_{cb}|}
\newcommand{\mupi}{\mu_\pi^2}
\newcommand{\mug}{\mu_G^2}
\newcommand{\rd}{\rho_D^3}
\newcommand{\rls}{\rho_{LS}^3}
\newcommand{\aspi}{\frac{\alpha_s}{\pi}}

\newcommand{\MS}{\ensuremath{\overline{\text{MS}}}}

\long\def\symbolfootnote[#1]#2{\begingroup%
\def\thefootnote{\fnsymbol{footnote}}\footnote[#1]{#2}\endgroup}

\def \be{\begin{equation}}
\def \ee{\end{equation}}
\newcommand{\bea}{\begin{eqnarray}}
\newcommand{\eea}{\end{eqnarray}}
\def \nn{\nonumber}

\begin{document}
%{\raggedright{\hspace{6cm} MITP/16-056}}

\title{Taming the higher power corrections in semileptonic $B$ decays}



%The names of authors may be listed in any order in the byline between the title and abstract. If the number of authors exceeds 40, the authors will be listed in the Table of Contents as, e.g., A. Jones et al. The author who submits the paper should ensure that all persons listed as authors approve the inclusion of their names, and check that the form of each name is the one normally used by that author.
%Byline addresses are set directly under the author names. They are intended to indicate the institution where the research was done. These addresses usually consist of department or division, institution, city, state or country. Street addresses, post office boxes, etc., may be included; zip and postal codes are proper.

\author{Paolo Gambino$^a$}
\author{Kristopher J. Healey$^a$}
\author{Sascha Turczyk$^b$}
%\thanks{}
\affiliation{$\,^a$ Dip. di Fisica, Universit\`a di Torino \& INFN, Torino, 10125 Torino, Italy}
\affiliation{$\,^b$ PRISMA Cluster of Excellence \& Mainz Institute for Theoretical Physics, Johannes Gutenberg University, 55099 Mainz, Germany}

\date{\today}
%There must be an abstract of no more than 600 characters, including spaces, which should be self-contained (no footnotes) for use in abstracting journals and databases. 
\begin{abstract}
We study the effect of dimension 7 and 8 operators on inclusive semileptonic $B$ decays and the extraction of $|V_{cb}|$.
Using moments of semileptonic $B$ decay spectra and information based on
the Lowest-Lying State saturation Approximation (LLSA)
we  perform a global fit of the
non-perturbative parameters of the heavy quark expansion including 
for the first time the $\mathcal{O}(1/m_b^{4,5})$ contributions.
Higher power corrections appear to have a very small effect on the extraction of $|V_{cb}|$, independently of the weight we attribute to 
the LLSA. 
%We obtain $|V_{cb}|  = (42.00 \pm 0.65)\times 10^{-3}$. %and $m_b^{kin} = 4.546(21)$GeV. 
 
 \end{abstract}

%\pacs{13.25.Hw, 14.40.Nd, 12.15.Hh, 12.15.Ff}
%\keywords{}

\maketitle
\section{Introduction} 
The results of the B Factories and LHC place stringent constraints on new physics
in the flavour sector. 
Only small deviations from the SM are allowed, and their detection represents
an experimental and theoretical challenge. In the next few years a wealth of new
experimental results will come from Belle-II and from the high-luminosity phase of
LHC. In this context, the precise determination of the parameters of the Cabibbo-
Kobayashi-Maskawa (CKM) matrix remains a high priority, as it is instrumental to
constraining new physics models and to setting bounds on the scale of new effective interactions. However, the determination of the
CKM element  $V_{cb}$, which  plays a special role in tests of the CKM unitarity and in FCNC transitions, is plagued by a long-standing 
$\sim\!\!3\sigma$ tension between the analyses based on inclusive and exclusive decays. This is unlikely 
to signal new physics \cite{Crivellin:2014zpa}  
and calls for a thorough investigation of all possible sources of theoretical uncertainty.

The  determination of $|V_{cb}|$
from inclusive semileptonic $B$ decays 
is based on an Operator Product Expansion (OPE) \cite{Chay:1990da,Bigi:1992su, Blok:1993va,Manohar:1993qn} which allows us to parameterize all of the non-perturbative physics in terms of the  expectation values of local operators in the $B$-meson
to be extracted from experimental data.
Since the contribution of higher dimensional operators is suppressed by powers of the 
heavy quark mass, only the operators of low dimension are expected to be relevant.
Current fits of inclusive semileptonic $B$ decays \cite{Alberti:2014yda} use experimental data on the moments of kinematic distributions to constrain the power corrections 
up to $1/m_b^3$ terms, corresponding to dimension $\le6$ operators, and neglect higher power corrections altogether. 

While present data appear to be  well described by these fits,  
investigations of higher power corrections 
are mandatory to test
the convergence of the heavy quark mass expansion.
 Moreover, the OPE does not lead to an expansion of inclusive observables 
 in inverse powers of $m_b$ 
 but also contains terms of ${\cal O}(1/m_b^n\ 1/m_c^k)$, with odd $n\ge 3$ and even $k\ge2$, sometimes dubbed {\it intrinsic charm} (IC) contributions \cite{Bigi:2005bh,Breidenbach:2008ua,Bigi:2009ym}, which alter the actual 
power counting since numerically $m_c^2\sim \Lambda_{QCD} m_b$
and thus ${\cal O}(1/m_b^3 m_c^2) \simeq {\cal O}(1/m_b^4)$. 
Higher power corrections have been 
studied in \cite{Dassinger:2006md,Mannel:2010wj}, where  nine new operators of  dimension 7 and 
eighteen new operators of dimension 8 have been identified and their Wilson coefficients computed at the tree-level. A rough estimate of 
the matrix elements of these 27 new operators is given by the 
Lowest-Lying State Approximation (LLSA)  \cite{Mannel:2010wj,Heinonen:2014dxa}, which 
assumes that the lowest lying heavy meson states saturate a sum-rule for the insertion of a heavy meson state sum. The LLSA 
relates higher-order matrix elements to lower dimensional ones and to the excitation energy $\epsilon$ and is expected to be valid within 50-100\% \cite{Heinonen:2014dxa}.

In this Letter, after  briefly reviewing the structure of the $1/m_b^{4,5}$  corrections
computed in \cite{Mannel:2010wj}, we study their inclusion in the fit of Ref.~\cite{Alberti:2014yda} and discuss how the results depend on the uncertainty associated to the LLSA.

\section{Power Corrections and Matrix Elements}
\label{sec:llsa}
Our analysis is based on the calculation of higher power corrections of  \cite{Mannel:2010wj}, 
which is performed at leading order in $\as$.
The  inclusive observables considered below (width, moments of kinematic distributions) can be calculated by an appropriate (weighted) phase-space integral of the differential decay width 
\be
  \text{d}\Gamma = 16 \pi G_F^2 |V_{cb}|^2 W_{\mu \nu} L^{\mu \nu}
\text{d}\phi\,,\label{Eq:dGamma}
\ee
where all the soft hadronic information is contained in the hadronic tensor $W_{\mu \nu} = -\frac{1}{\pi}\,\text{Im}\,T_{\mu \nu}$. The hadronic tensor is the imaginary part of
the forward matrix element of  a time-ordered product of weak currents.
The charm quark in this forward matrix element propagates in a background 
field. We expand the background field propagator
$S_\text{BGF}$, with momentum $Q^\mu= m_b v^\mu + k^\mu - q^\mu$,   in powers of $k^\mu/m_b$, where  $k^\mu \to iD^\mu$ is 
the residual momentum of the $b$-quark inside the $B$-meson
\begin{align} 
	&T_{\mu \nu} = \langle B(p) | \bar{b}_v \Gamma_\mu i S_{\rm BGF}  \Gamma^\dagger_\nu b_v | B(p) \rangle \nonumber \\&=  \sum_i  {\rm Tr}  \left\{ \Gamma_\mu  \frac{1}{\slashed{Q}-m_c+i \epsilon}  \Gamma^\dagger_\nu  \, \hat{\Gamma}^{(i)} \right\}   A^{(i,0)} \nonumber \\ 
	&+  \sum_i  {\rm Tr}  \left\{ \Gamma_\mu  \frac{1}{\slashed{Q}-m_c+i \epsilon} \gamma^{\mu_1}  \frac{1}{\slashed{Q}-m_c+i \epsilon} \Gamma^\dagger_\nu   \, \hat{\Gamma}^{(i)}  \right\}  A^{(i,1)}_{\mu_1}  \nonumber \\
	&+ \cdots   \label{Eq:Trace}
\end{align}
The coefficients  $A^{(i,m)}_{\mu_1 \mu_2\ldots\mu_m}$ containing the non-perturbative parameters are known analytically at ${\cal O} (1/m_b^2)$ \cite{Blok:1993va,Manohar:1993qn} (corresponding to $m=2$),  at ${\cal O} (1/m_b^3)$ \cite{Gremm:1996df}, and at order $1/m_b^{4,5}$~\cite{Mannel:2010wj}. At the lowest non-trivial order, corresponding to dimension 5 operators,  the non-perturbative parameters are given by
\bea
    2M_B\, \mu_\pi^2 &=& - \langle \bar B| \bar b_v \, i D_\rho i D_\sigma  \,b_v |\bar B\rangle \,\,  \Pi^{\rho \sigma},   \label{Eq:dim5Spinless} \\
    2M_B\, \mu_G^2 &=& \frac12 \langle \bar B|  \bar b_v \,\big[ i D_\rho , iD_\sigma\big]  \big (-i \sigma_{\alpha \beta}\big)\,b_v  |\bar B \rangle  \,\,  \Pi^{\alpha \rho} \Pi^{\beta \sigma} \label{Eq:dim5Spin}\nn \,,
\eea
where $\Pi^{\mu\nu}=g^{\mu\nu}-v^\mu v^\nu$, and $v^\mu$ is the heavy quark velocity.
At each higher order in $1/m_b$ we have one more derivative in $A^{(i,m)}_{\mu_1 \mu_2\ldots\mu_m}$. Thus the number of parameters proliferates. We have only 2 parameters, $\rd$ and $\rls$, at $O(1/m_b^3)$, but there are nine  additional ones at $O(1/m_b^4)$ and eighteen at $O(1/m_b^5)$. As mentioned in the Introduction, upon integration over the phase space the Wilson coefficient of some of the dimension 8 operators are sensitive to the (infrared)
charm mass scale and represent the IC terms of $O(1/m_b^3 m_c^2)$, which numerically dominate the  $O(1/m_b^5)$ contributions.

In the following we will include the $O(1/m_b^{4,5})$ corrections in the 
fit to the semileptonic moments on which the inclusive determination of $|V_{cb}|$ is based.
We will use the LLSA ansatz, proposed in \cite{Mannel:2010wj} and made  more systematic 
in \cite{Heinonen:2014dxa}, to constrain the 27 new parameters.

The goal of LLSA is to estimate expectation values of local operators of the form $\bar b_v iD_{\mu_1} iD_{\mu_2} \ldots iD_{\mu_n}\Gamma b_v$, where $\Gamma$ is
a Dirac matrix. Splitting the chain of covariant derivatives
into two shorter ones labeled by $A_1^k$ and $C_k^n$ and inserting 
a full set of  intermediate states between them one finds in the heavy quark limit \cite{Mannel:2010wj, Heinonen:2014dxa} 
\bea
&&     \langle \bar B |  \bar{b}_v  A_1^k \, C_k^n \,\Gamma \, b_v| \bar B \rangle  = \label{Eq:Decomp}\\
&&\qquad\frac{1}{2M_B} \sum_n \, 
    \langle \bar B |  \bar{b}_v \, A_1^k \,b_v(0) | H_n \rangle    \langle H_n | \bar{b}_v(0) \, C_k^n \,\Gamma \, b_v | \bar B \rangle \,, \nn
\eea
where $| H_n \rangle $ are hadronic states 
with the appropriate quantum numbers.
The LLSA assumes that the sum of intermediate states is saturated by the 
lowest-lying state that can contribute, {\it i.e.}\ either the ground-state multiplet $B, B^*$ or the 
first excited states with $\ell=1$. %the same content of light degrees of freedom. 
Indeed, the matrix elements involving time derivatives like  $\langle B| \bar{b} iD_j iD_0^k iD_l %[\sigma] 
b | B \rangle$ are saturated by $P$-wave intermediate states, with parity opposite to that of the ground state. Including these states in the sum leads to extra powers of the $P$-wave excitation energy, ${\epsilon} = M_P - M_B$. While there exist separate contributions coming from the spin $\frac12, \frac32$ light degrees of freedom, we assume ${\epsilon}_{1/2} = {\epsilon}_{3/2} = {\epsilon} \simeq 0.4 \mbox{GeV}$. 

\begin{scriptsize}
\begin{table}[t]
\begin{equation} \nonumber
\begin{array}{|lc|lc|}\hline\hline
   \overline{m}_{1} & \frac{5 (\mu^2_{\pi})^{2}}{9} & r_{6} &  \epsilon^2 \rd \\
 \overline{m}_{2} &  -\epsilon \rd & r_{7} &  0 \\
 \overline{m}_{3} &  -\frac{(\mu^2_G)^2}{6} & r_{8} &  \epsilon^2 \rls \\
 \overline{m}_{4} &  \frac{(\mu^2_G)^2}{8}+\frac{(\mu^2_{\pi})^2}{6} & r_{9} &  -\mu^2_{\pi} \rls \\
 \overline{m}_{5} &  -\epsilon \rls & r_{10} &  \mu^2_G \rd \\
 \overline{m}_{6} &  \frac{(\mu^2_G)^2}{6} & r_{11} & {\scriptscriptstyle \frac{\mu^2_G \rd}{3}-\frac{\mu^2_G \rls}{6}+\frac{\mu^2_{\pi} \rls}{3}} \\
 \overline{m}_{7} &  -\frac{\mu^2_G \mu^2_{\pi}}{3} & r_{12} &  {\scriptscriptstyle -\frac{\mu^2_G \rd}{3}-\frac{\mu^2_G \rls}{6}-\frac{\mu^2_{\pi} \rls}{3} }\\
 \overline{m}_{8} &  -\mu^2_G \mu^2_{\pi} &
  r_{13} &{\scriptscriptstyle  -\frac{\mu^2_G \rd}{3}+\frac{\mu^2_G \rls}{6}+\frac{\mu^2_{\pi} \rls}{3}} \\
 \overline{m}_{9} &  \frac{(\mu^2_G)^2}{8}-\frac{5 \mu^2_G \mu^2_{\pi}}{12} &
r_{14} & {\scriptscriptstyle
\rls \left(\epsilon^2 +\frac{\mu^2_G}{6}-\frac{\mu^2_{\pi}}{3} \right)+\frac{\mu^2_G \rd}{3} } \\
 r_{1} &  \epsilon^2 \rd & r_{15} &  0 \\
 r_{2} &  -\mu^2_{\pi} \rd & r_{16} &  0 \\
 r_{3} &  -\frac{\mu^2_G \rls}{6}-\frac{\mu^2_{\pi} \rd}{3} & r_{17} &  \epsilon^2 \rls \\
 r_{4} &  {\scriptscriptstyle \epsilon^2 \rd+\frac{\mu^2_G \rls}{6}-\frac{\mu^2_{\pi} \rd}{3}}& r_{18} &  0 \\
 r_{5} &  0 & \, & \\ \hline\hline
\end{array}
\end{equation}
\caption{LLSA expressions  for the higher-order non-perturbative parameters.}
\label{table:llsa}
\end{table}
\end{scriptsize}

In the following we use the notation of \cite{Mannel:2010wj}, according to which 
the nine matrix elements that occur at $O(1/m^4)$ are denoted by $m_i$, and the eighteen
at $O(1/m^5)$ by $r_i$. The operators involved coincide with those 
identified in \cite{Heinonen:2014dxa}, 
% Note that Refs.~\cite{Mannel:2010wj} and \cite{Heinonen:2014dxa} are consistent, 
 even though different notations are adopted.
% they look different because of  different notations. 
It is useful to 
%In the following we use the notation of \cite{Mannel:2010wj}, but we   
redefine the $1/m_b^4$ parameters to account for combinatorial factors.  
In practice, we expand the (anti-)commutators and count the number of terms
after expunging those which
are of higher order in $1/m_b$ due to the equations of motion.
 We then expect the parameters to have a {\it natural} scale of
 $O(\Lambda_{QCD}^n)$, with $n$ the dimension of the corresponding operator,
as  is also the case for the parameters in Eq.~(\ref{Eq:dim5Spinless}). 
  The rescaled parameters are %given by
\bea
\nn\overline{m}_1 &=&m_1 \ \  \ \ \qquad \overline{m}_2 = m_2 \qquad \ \ \  \overline{m}_3 = m_3/4 \\
\overline{m}_4 &=& m_4/8 \  \qquad \overline{m}_5 = m_5 \ \ \ \qquad  \overline{m}_6 = m_6/4 \\
\nn \overline{m}_7 &=& m_7/8  \ \qquad  \overline{m}_8 = m_8/8 \qquad \overline{m}_9 = m_9/8\,.
\eea
No such redefinition is necessary for the $1/m_b^5$ parameters, as they were already defined in this way. The LLSA expressions for the $\overline{m}_i,r_i$ are reported in Table~\ref{table:llsa}.

\section{Inclusive Observables}
The 
OPE allows us to express sufficiently inclusive observables as a double series in $\alpha_s$ and $\Lambda_{QCD}/m_b$. In fact, the non-perturbative corrections to the semileptonic differential rate start at  ${\cal O}(1/m_b^2)$.
Perturbative corrections are known up to NNLO~\cite{Pak:2008qt,Melnikov:2008qs,Biswas:2009rb,Gambino:2011cq} and the mixed ${\cal O}(\alpha_s \mu^2_{\pi,G}/m_b^2)$ corrections \cite{Alberti:2012dn,Alberti:2013kxa,Mannel:2015jka}  
 have also been calculated. The expansion requires knowledge of the expectation values of local operators in the $B$-meson. These non-pertubative parameters can be determined from measurements of the normalized moments of the lepton energy and invariant hadronic mass distributions in inclusive $B\to X_c\ell\nu$ decays,
\bea
  \langle
  E^n_\ell\rangle&=&\frac{1}{\Gamma_{E_\ell>E_\mathrm{cut}}}\int_{E_\ell>E_\mathrm{cut}}
   E^n_\ell\ \frac{d\Gamma}{dE_\ell} \ dE_\ell~,\\
  \langle
  M^{2n}_X\rangle&=&\frac{1}{\Gamma_{E_\ell>E_\mathrm{cut}}}\int_{E_\ell>E_\mathrm{cut}}
  M^{2n}_X\ \frac{d\Gamma}{dM^2_X}\ dM^2_X~,\nn
\eea
where $E_\ell$ is the lepton energy, $m_X^2$ the invariant hadronic mass squared and $E_{cut}$ an experimental lower cut on the lepton energy applied by the experiments. The cut dependence of the moments  provides additional information on the OPE parameters we are fitting. For moments with $n>1$,  it is convenient to employ {\it central} moments, computed relative to $\langle E_\ell\rangle\equiv \ell_1$ and $\langle m^2_X\rangle\equiv h_1$,
\bea
  \nonumber \ell_n(E_{cut}) &=& \langle (E_\ell - \langle E_\ell\rangle )^n\rangle_{E_\ell > E_{cut}},\\
    h_n(E_{cut}) &=& \langle (M_X^2 - \langle M_X^2\rangle )^n\rangle_{E_\ell > E_{cut}}.
\eea
We also have information on the lepton energy cut dependence of the inclusive width, which can be studied introducing $ R^{*}=\Gamma_{E_\ell>E_\mathrm{cut}} /\Gamma_{\mathrm{tot}}$.
The information on the non-perturbative parameters obtained from a fit to these observables enables us to then extract $|V_{cb}|$ from the total semileptonic width \cite{Gambino:2004qm,Aubert:2004aw,Bauer:2004ve,Gambino:2013rza,Alberti:2014yda}. 

All analyses have so far considered only the minimal set of four matrix elements
which appear at $\mathcal{O}(1/m_b^{2,3})$.
The $\mathcal{O}(1/m_b^{4,5})$ contributions have never been included, although 
a rough estimate of their importance has been given in \cite{Mannel:2010wj}.
From the results of that paper
we have computed all the $\mathcal{O}(1/m_b^{4,5})$ corrections to the first three hadronic and leptonic moments and to $R^*$; we will now employ 
these expressions in the global fit to determine $|V_{cb}|$. The result for the  width is given  in the Appendix. 
 Notice that 
normalized moments are ratios of two heavy quark expansions; re-expanding these ratios
in inverse powers of $m_b$ one finds  that the $\mathcal{O}(1/m_b^{4,5})$ corrections 
also include products of $\mathcal{O}(1/m_b^2)$ with $\mathcal{O}(1/m_b^{2,3})$ terms.

\section{The Fit}
We upgrade the fit strategy introduced in \cite{Gambino:2013rza} in the kinetic scheme, and use as a baseline the default parameters and settings most recently employed in \cite{Alberti:2014yda}. In particular, we use the same experimental data;
the full list of  available measurements  \cite{Aubert:2009qda,Aubert:2004td,Urquijo:2006wd,Schwanda:2006nf,Acosta:2005qh,Csorna:2004kp,Abdallah:2005cx}
and the leptonic energy cuts employed in the fit is given in Table 1
of Ref.~\cite{Gambino:2013rza}.
 We also employ the $\MS$ scheme for the charm mass and
use the constraints $\overline{m_c}({\rm 3GeV})=0.986(13)$GeV \cite{Chetyrkin:2009fv},
$\mug(m_b)=0.35(7)$GeV$^2$, $\rls=-0.15(10)$GeV$^3$.

The inclusion of higher power corrections allows us to slightly
decrease the theoretical errors, which are estimated using the method of
Ref.~\cite{Gambino:2013rza}, i.e.\ varying the HQE parameters by fixed amounts in the calculation of an observable. 
Here we use the same settings as in \cite{Alberti:2014yda}, except for the variation 
in $\rho^3_{D,LS}$, which we decrease from 30\% to 22\%, to take into account the inclusion
of ${\cal O}(1/m^4,5)$ power corrections.
For what concerns the correlations among theoretical errors 
we choose scenario $\mathbf D$ of Ref.~\cite{Gambino:2013rza}, 
where different central moments are uncorrelated and 
the correlation between measurements of the same moment with $E_{cut}$ differing by 100\,MeV is given by a factor which becomes smaller for increasing $E_{cut}$. 
%\begin{small}
%\begin{table}[t]
%\begin{tabular}{|c||c|c|c|}
%\hline
%Experiment  &   & $E_{(\ell)}^{cut} (\mbox{GeV})$ & Ref.   \\ \hline
%Babar & $R^{*}$  &  0.6, 1.2, 1.5 & \cite{Aubert:2009qda,Aubert:2004td} \\
%Babar & $\ell_{1}$&  0.6, 0.8, 1, 1.2, 1.5 & \cite{Aubert:2009qda,Aubert:2004td} \\ 
%Babar & $\ell_{2}$&  0.6, 1.0, 1.5 & \cite{Aubert:2009qda,Aubert:2004td}\\ 
%Babar & $\ell_{3}$&  0.8, 1.2 & \cite{Aubert:2009qda,Aubert:2004td}\\ 
%Babar & $h_{1}$& 0.9, 1.1, 1.3, 1.5 & \cite{Aubert:2009qda} \\ 
%Babar & $h_{2}$&  0.8, 1.0, 1.2, 1.4 & \cite{Aubert:2009qda} \\ 
%Babar & $h_{3}$&   0.9, 1.3  & \cite{Aubert:2009qda} \\ 
%Belle & $R^{*}$  &  0.6, 1.4 & \cite{Urquijo:2006wd}  \\
%Belle & $\ell_{1}$&  1.0, 1.4 & \cite{Urquijo:2006wd} \\ 
%Belle & $\ell_{2}$&  0.6, 1.4 & \cite{Urquijo:2006wd} \\ 
%Belle & $\ell_{3}$& 0.8, 1.2 & \cite{Urquijo:2006wd} \\ 
%Belle & $h_{1}$&  0.7, 1.1, 1.3, 1.5 & \cite{Schwanda:2006nf} \\ 
%Belle & $h_{2}$&  0.7, 0.9, 1.3 & \cite{Schwanda:2006nf} \\  
%CDF & $h_{1,2}$&  0.7 & \cite{Acosta:2005qh} \\  
%CLEO & $h_{1,2}$&  1.0, 1.5 & \cite{Csorna:2004kp} \\  
%DELPHI & $\ell_{1,2,3}$&  0.0 & \cite{Abdallah:2005cx} \\  
%DELPHI & $h_{1,2,3}$&  0.0 & \cite{Abdallah:2005cx} \\  
%        \hline
%\end{tabular}
%\caption{Experimental Data used in the fit and their associated lower leptonic energy cut.}
%\label{table:expdata}
%\end{table}
%\end{small}

The results of the default fit performed in \cite{Alberti:2014yda} read
\begin{eqnarray}\nonumber
 m_b^{kin} = 4.553(20), & \quad \overline{m}_c(3\mbox{GeV}) &= 0.987(13), \\
 \mu^2_{\pi} = 0.465(68),  &   \quad\qquad\mu^2_G &= 0.332(62), 
  \label{def2014}
  \\ \nonumber \rho_D = 0.170(38), & \quad\qquad\rls &= -0.150(96),
 \end{eqnarray}
where all parameters except for $m_c$ are in the kinetic scheme with cutoff $\mu_{kin}=1$GeV. Using $\tau_B=1.579$ps, Ref.~\cite{Alberti:2014yda} gets $|V_{cb}|=42.21(78)\ 10^{-3}$ .

As a first step in the analysis, we repeat exactly the same fit to 
the $\mathcal{O}(1/m_b^{2,3})$ parameters
but include the 
$\mathcal{O}(1/m_b^{4,5})$ corrections in the theoretical predictions. We fix their
values using the LLSA expressions for the matrix elements $\overline{m}_i,r_i$, computed 
using the central values in (\ref{def2014})
and $\epsilon=0.4$GeV. The products
 of $1/m_b^2$ and $1/m_b^3$ effects are also computed using (\ref{def2014}) and cannot vary in the fit.
The results  are similar to those in (\ref{def2014}), except that $\mupi$ and $\rd$ get a significant shift up, 
$\mupi=0.506(74)$GeV$^2$, $\rd=0.257(42)$GeV$^3$, and that the central value of  $ |V_{cb} |$ is $ 42.47 \ 10^{-3}$.
This total 0.7\% increase in $ |V_{cb} |$ occurs despite the $\mathcal{O}(1/m_b^{4,5})$ contributions increase the semileptonic width by more than 1\%, leading to a {\it direct} reduction of $ |V_{cb} |$. A similar pattern (larger $\mupi,\rd$, and $ |V_{cb} |$) is observed
if we fix only the matrix elements $\overline{m}_i,r_i$ to their LLSA values, and let
the products of $1/m_b^2$ and $1/m_b^3$ effects to vary.
\begin{table}[t]
\begin{small}
\begin{equation}\nonumber
\begin{array}{|lrr|lrr|} \hline
 m_b^{kin} &  4.546  &  0.021  			&  r_1  &  0.032  &  0.024  \\
 \overline{m}_c(3{\rm GeV})  & 0.987 & 0.013 	&  r_2  & -0.063 & 0.037 \\
 \mu^2_{\pi}  & 0.432 & 0.068			&  r_3  & -0.017 & 0.025 \\
 \mu^2_{G}  & 0.355 & 0.0	60			&  r_4  & -0.002 & 0.025 \\
 \rho^3_{D}  & 0.145 & 0.061 			&  r_5  & 0.001 & 0.025 \\
 \rho^3_{LS}  & -0.169 & 0.097 			&  r_6  & 0.016 & 0.025 \\
 \overline{m}_{1}  & 0.084 & 0.059 		&  r_7  & 0.002 & 0.025 \\
 \overline{m}_{2}   & -0.019 & 0.036	&  r_8  & -0.026 & 0.025 \\
 \overline{m}_{3}   & -0.011 & 0.045 	&  r_9  & 0.072 & 0.044 \\
 \overline{m}_{4}   & 0.048 & 0.043 	&  r_{10}  & 0.043 & 0.030 \\
 \overline{m}_{5}   & 0.072 & 0.045 	&  r_{11}  & 0.003 & 0.025 \\
 \overline{m}_{6}   & 0.015 & 0.041 	&  r_{12}  & 0.018 & 0.025 \\
 \overline{m}_{7}   & -0.059 & 0.043 	&  r_{13}  & -0.052 & 0.031 \\
 \overline{m}_{8}   & -0.178 & 0.073 	&  r_{14}  & 0.003 & 0.025 \\
 \overline{m}_{9}   & -0.035 & 0.044 	&  r_{15}  & 0.001 & 0.025 \\
 \chi^2/dof 		& 0.46 &  			&  r_{16}  & 0.001 & 0.025 \\
 BR(\%)  & 10.652 & 0.156 				&  r_{17}  & -0.028 & 0.025 \\
 \mathbf{10^3 |V_{cb}|}  & \mathbf{42.11} & \mathbf{0.74} &  r_{18}  & -0.001 & 0.025	\\ \hline
\end{array}
\end{equation}
\end{small}
\caption{Default fit results: the second and third columns give the central values and standard deviations.}
\label{table:fitresults}
\end{table}

While the LLSA can set the scale of the higher power effects, it is certainly subject to 
large corrections. We therefore assign an error to the LLSA predictions and assume gaussian priors for all the $\overline{m}_i,r_i$, which are then fit along with the other parameters.
The accuracy of the LLSA  is hard to quantify. 
 At ${\cal O}(1/m_b^3)$  the values of 
$\rho_D^3$ and $\rho_{LS}^3$ in (8) match well the LLSA expressions $\rho_D^3=\epsilon \mu_\pi^2$
and $\rho_{LS}^3=-\epsilon \mu_G^2$. 
Ref.~\cite{Heinonen:2014dxa} estimates a $\sim50$\% uncertainty, which obviously does not hold
when the LLSA leads to zero matrix elements. 
Ref.~\cite{Gambino:2012rd} in Sec.~6.5 found indications for large non-factorizable 
corrections, which could reach 100\% in some expectation values not affected by cancellations.
Dimensionally, we know that
the non-perturbative parameters of the OPE are quantities of ${\cal O}(\Lambda_{QCD}^n)$. 
There are in fact two scales involved in their determination: $M_B-m_b$ and the 
mass splitting $\epsilon\simeq 0.4$~GeV between the $B$ meson and the lowest $P$-wave excitation. 
Accordingly, we prescribe the error to be the maximum of either $60\%$ of the parameter's value or $\Lambda_{LL}^n/2$ ($n=4,5$), where we use a scale $\Lambda_{LL} = 0.55 $~GeV
which roughly corresponds to the   average of the two relevant scales.
The fit is performed starting with LLSA central values based on %the results of the 2014 fit,
 Eq.~(\ref{def2014}) and $\epsilon=0.4$GeV. The LLSA central values are then updated to the results of the new fit,
iterating the procedure until the results stabilize. 
 \begin{figure}[t]
 \begin{tabular}{cc}
\begin{adjustbox}{valign=t}
\begin{tabular}{@{}c@{}}
  \includegraphics[width=2.3in]{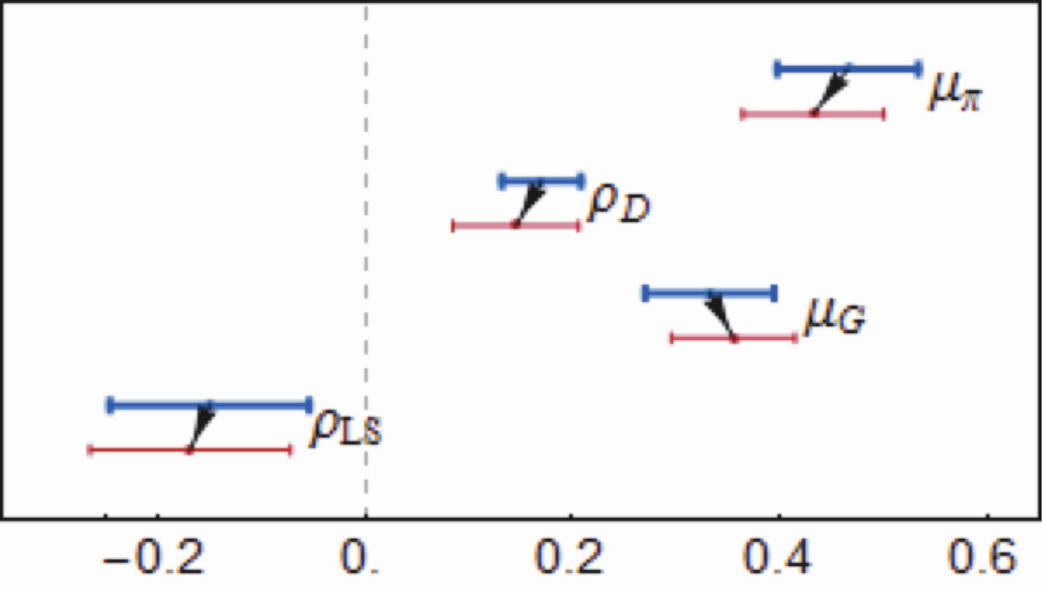}\\[-30px]
  \includegraphics[width=2.3in]{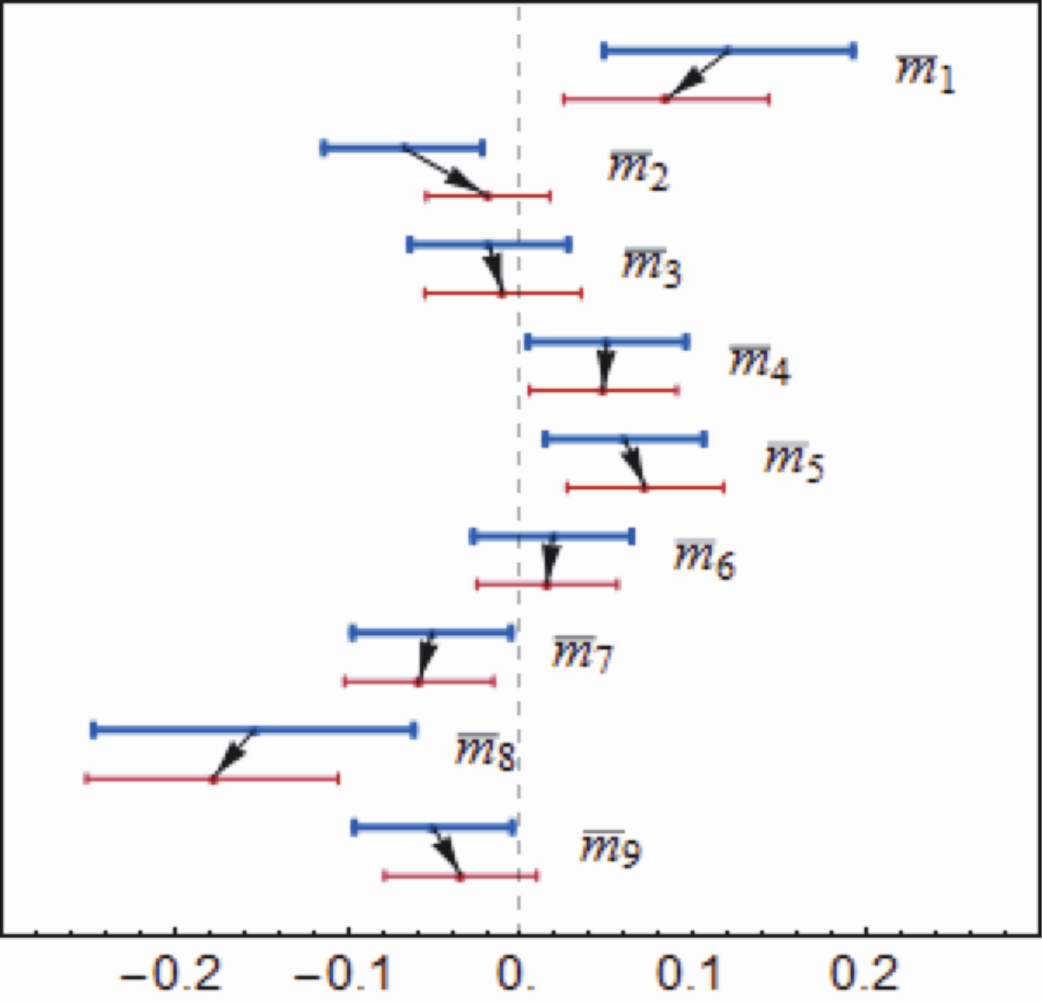}
\end{tabular}
\end{adjustbox}
&
\begin{adjustbox}{valign=t}
  \includegraphics[width=2.3in]{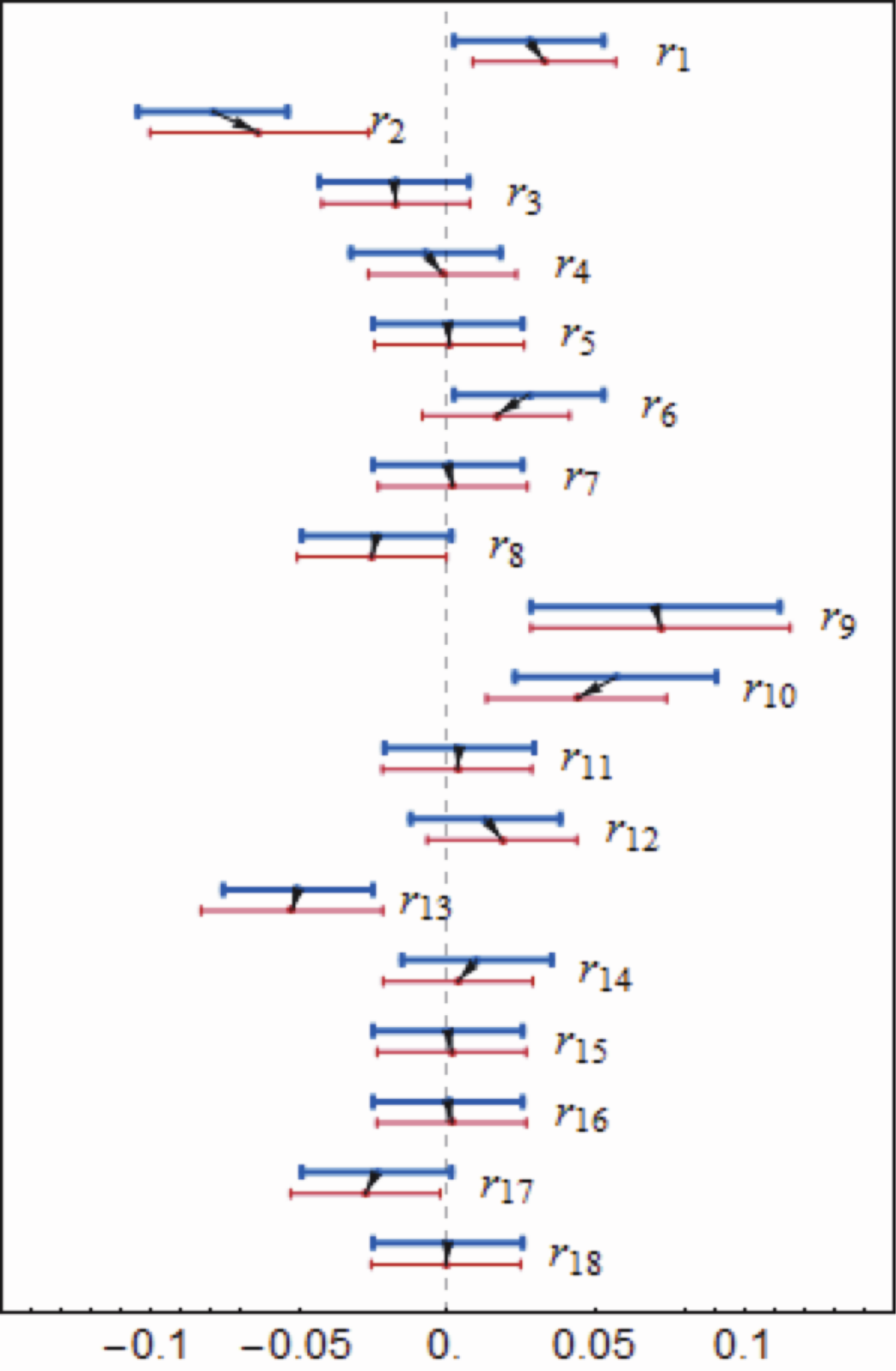}
\end{adjustbox}
\end{tabular}
 \caption{Shifts in the OPE parameters from the LLSA using the 2014 fit (blue thick) results to the current fit including higher-order corrections (red thin). Error bars represent the error in the priors and the resulting fit error, respectively.}
\label{fig:fitshiftplots}
\end{figure}

 \section{Results}
We report the results of the default fit in Table \ref{table:fitresults}. 
In Fig.~\ref{fig:fitshiftplots} we compare the $\mu^2_{\pi,G},\rho^3_{D,LS}$ results of the 2014 fit in (\ref{def2014}) with those of the  new default fit. We also 
compare the LLSA predictions for $\overline{m}_i, r_i$ based on (\ref{def2014})  with the results of the default fit. The LLSA uncertainty is computed as explained in the previous paragraph.
 We can see that most of the new parameters do not change much from their LLSA value, 
reflecting the low sensitivity of the fit to  higher power parameters. However, there are exceptions, especially among the $\overline{m}_i$:
 the largest shift occurs for $\overline{m}_2$ and corresponds to $1.2\sigma_{LLSA}$. Indeed, the hadronic moments at higher cuts are specifically sensitive to some of the
 $\overline{m}_i$, see Eqs.(\ref{dhi}) in the Appendix. 
Using the fit results we compute the total semileptonic width, also reported
in the Appendix, and comparing it to the BR in Table II divided by $\tau_B$, we get $\vcb$.  
 The value of $|V_{cb}|$ is remarkably close to that obtained in \cite{Alberti:2014yda} and
 the quality of the fit is very good, $\chi^2/dof=0.46$, but somewhat higher than in \cite{Alberti:2014yda}. 
 \begin{figure}[t]
\begin{tabular}{cc}
  \includegraphics[width=2in]{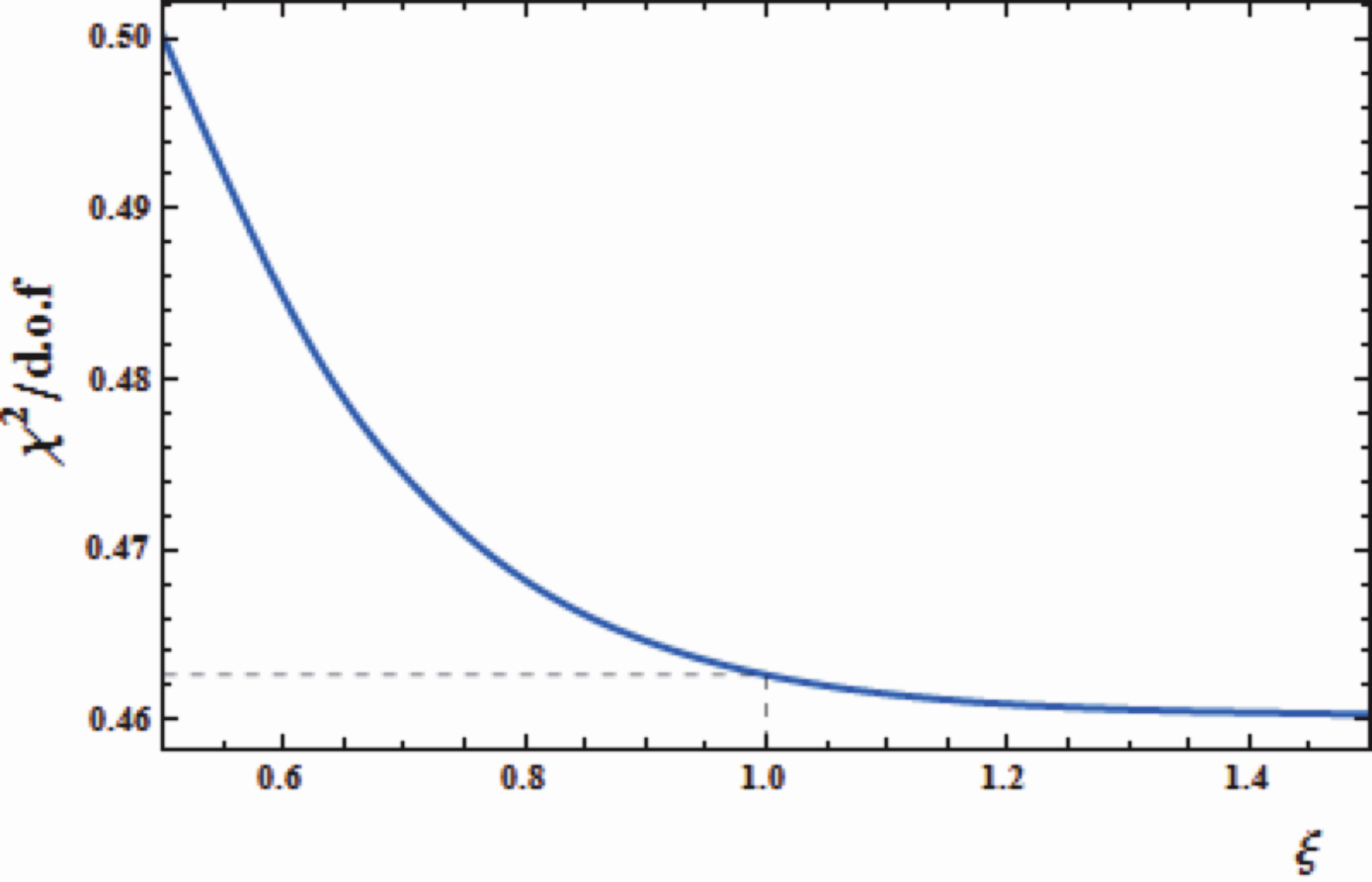} &   \includegraphics[width=2in]{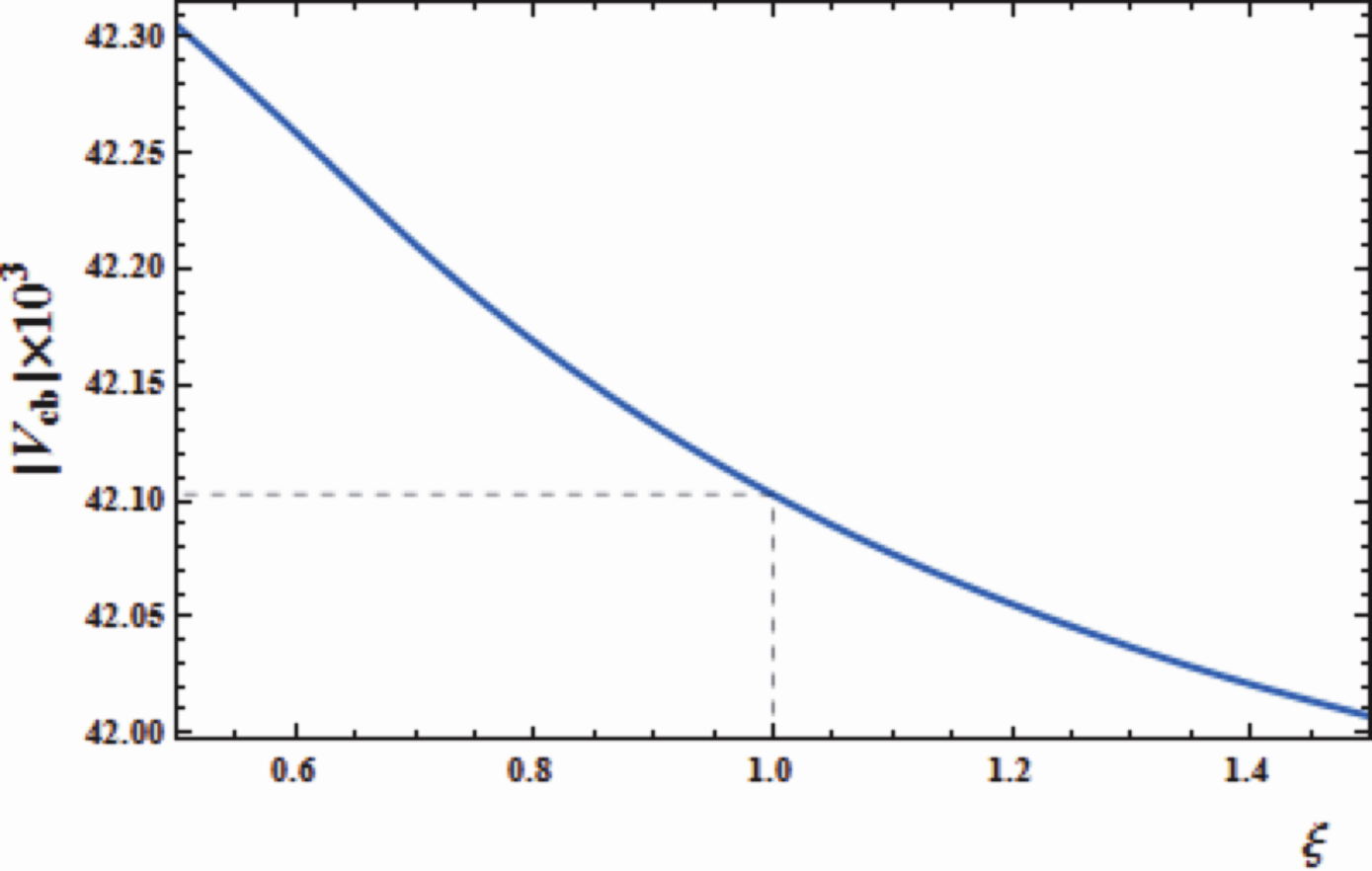} \\
 \includegraphics[width=2in]{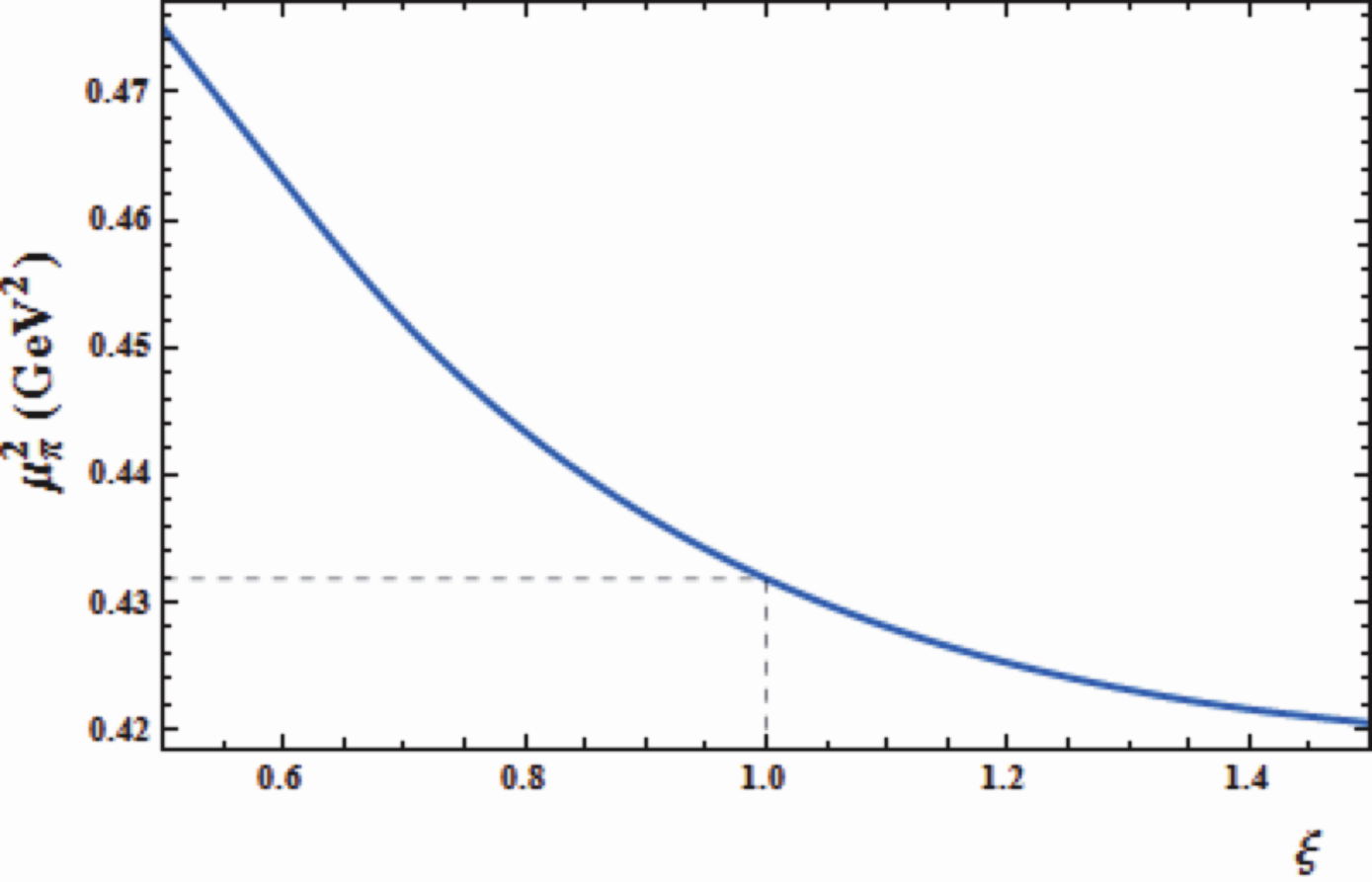} &   \includegraphics[width=2in]{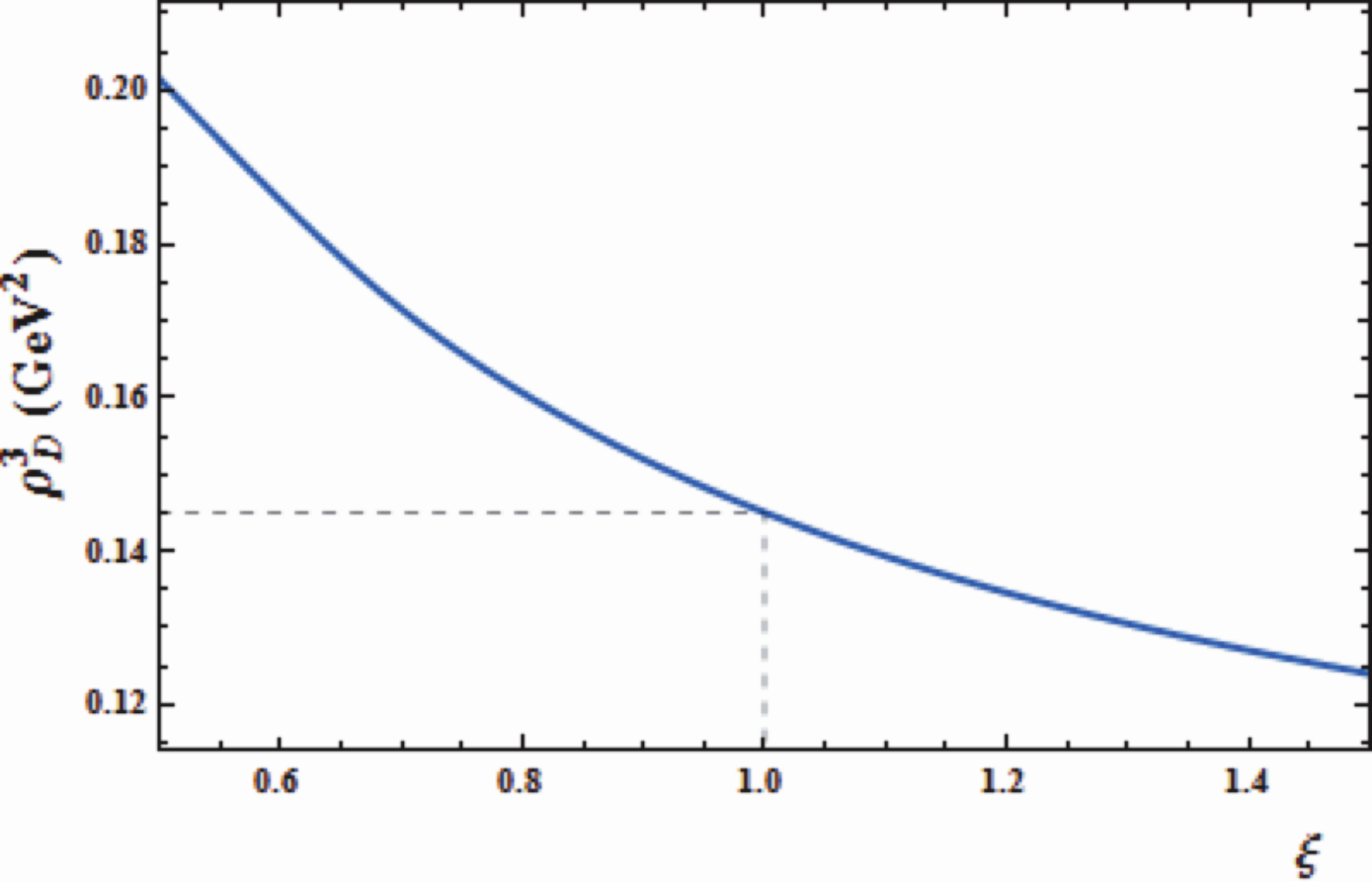}  
\end{tabular}
\caption{Dependence of the fit results as a function of the LLSA uncertainty. }
\label{fig:uncertplots}
\end{figure} 

\begin{figure}[t]
\begin{tabular}{cc}
  \includegraphics[width=2in]{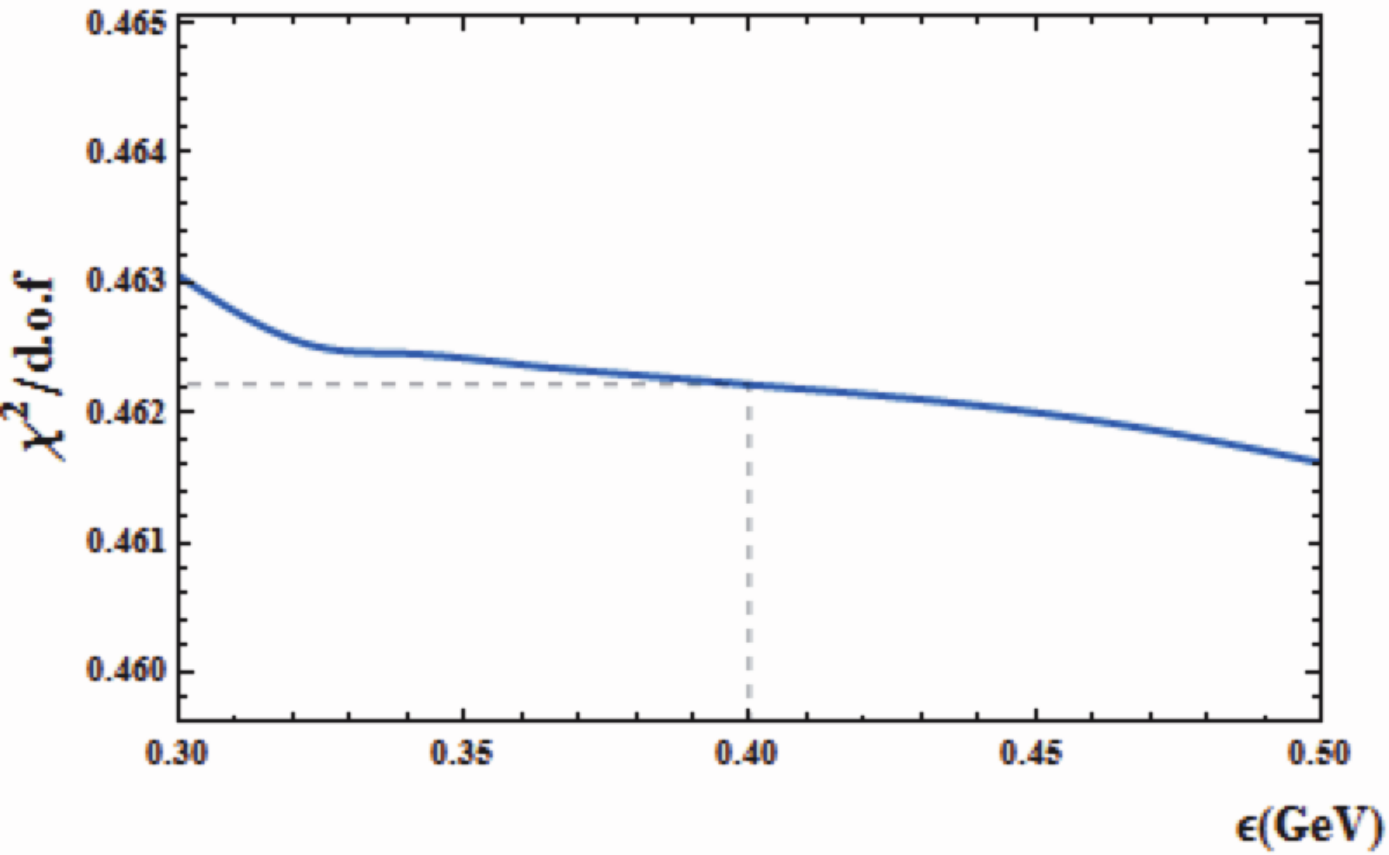} &   \includegraphics[width=2in]{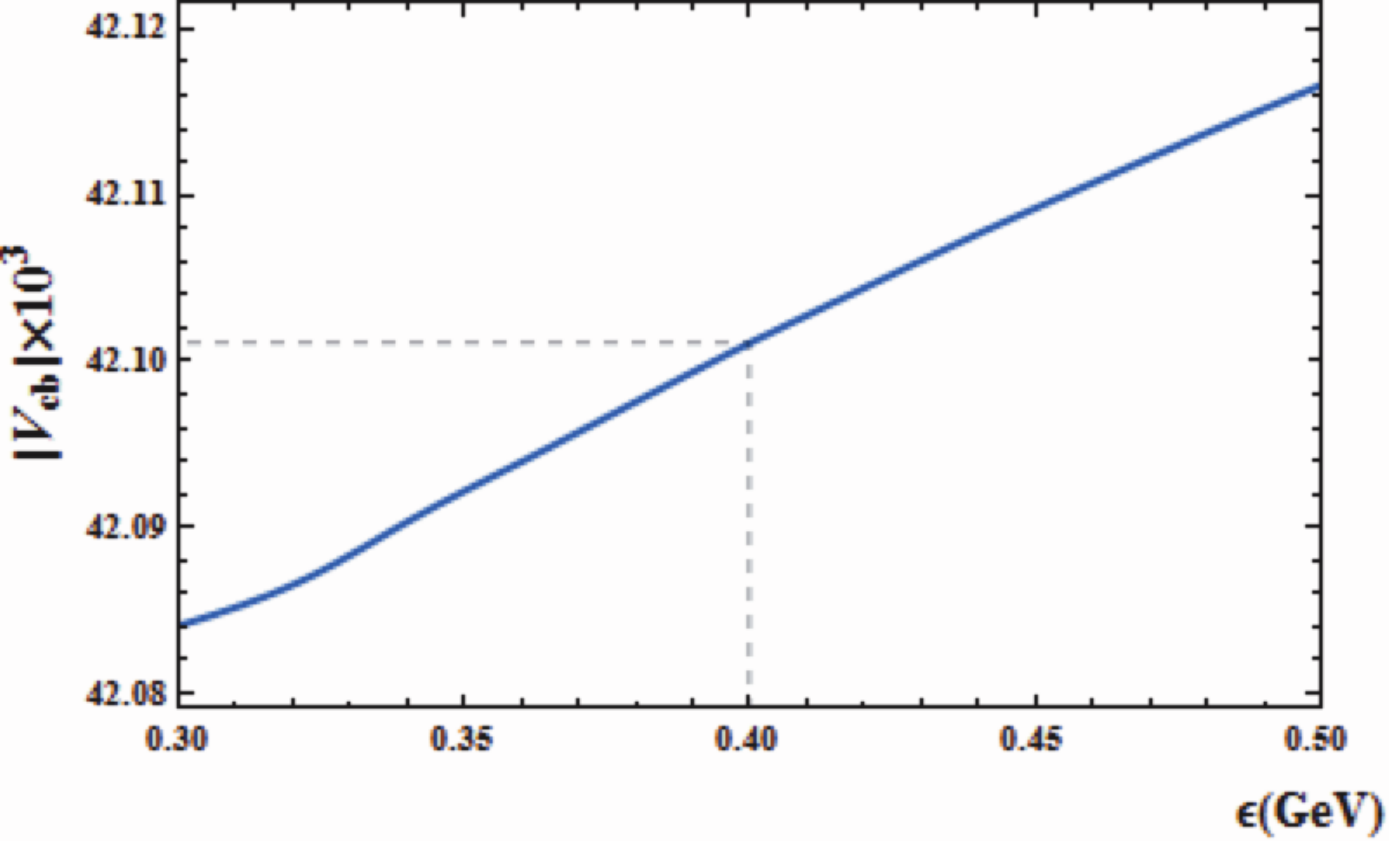} \\
 \includegraphics[width=2in]{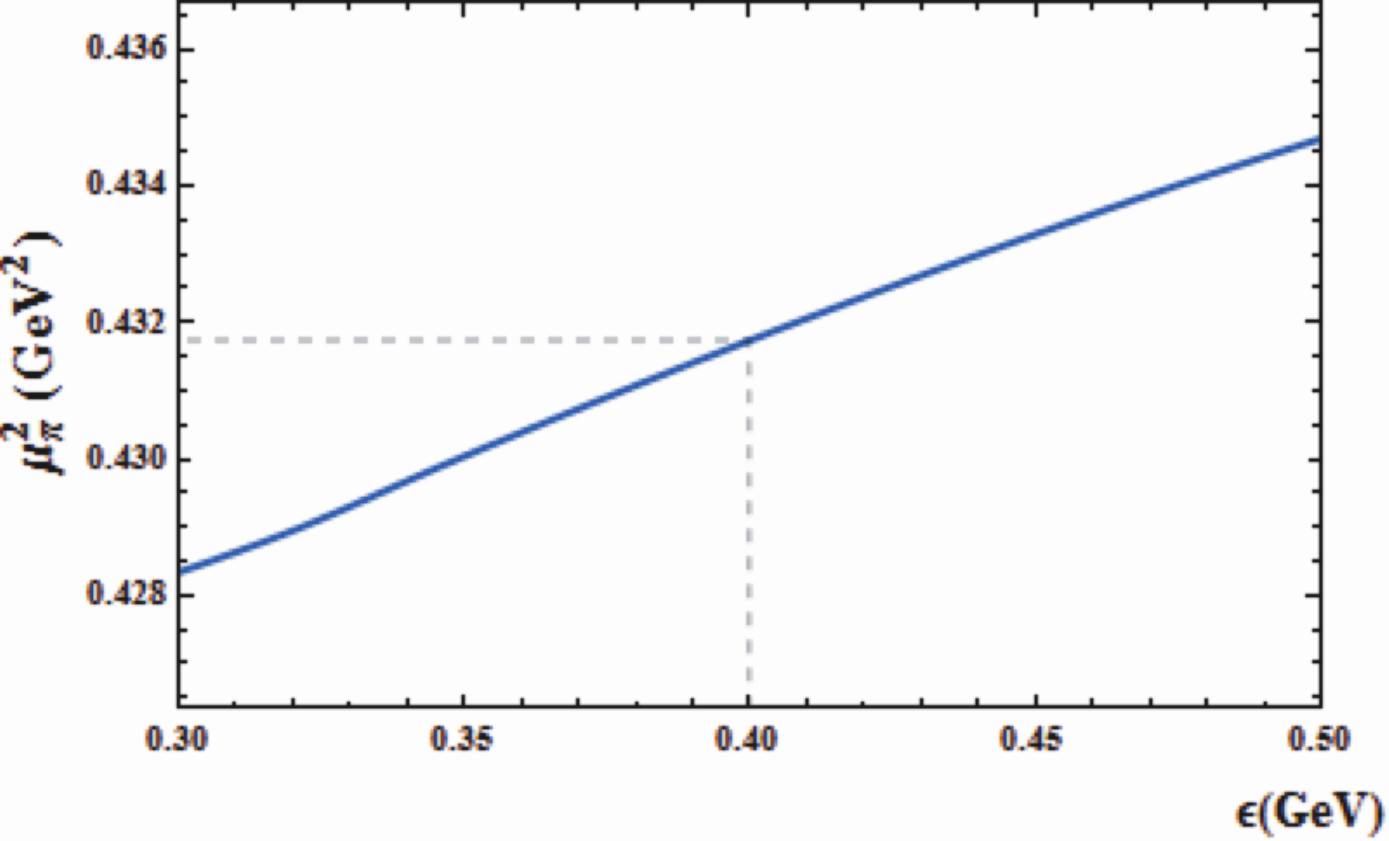} &   \includegraphics[width=2in]{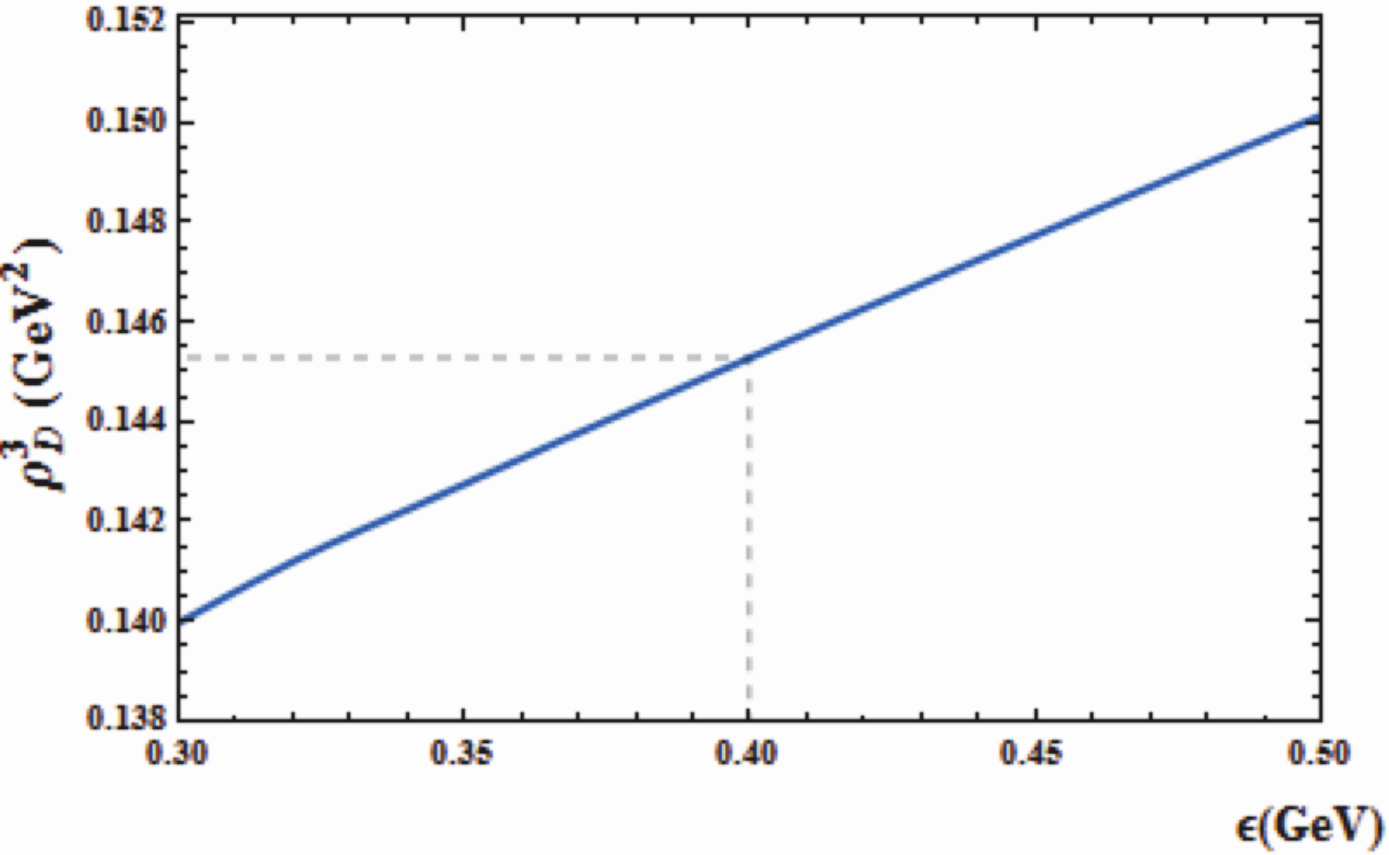}  
\end{tabular}
\caption{
Dependence of the fit results as a function of the P-wave excitation energy ${\epsilon}$.} \label{fig:epsilonplots}
\end{figure}

To verify the stability of the fit with respect to the choices we made for the LLSA uncertainty, 
we varied this uncertainty by a  multiplicative factor $\xi$.
The results are 
shown in Fig.~\ref{fig:uncertplots}: $|V_{cb}|$ changes very little. 
Of course, increasing the uncertainty on  the higher-order matrix elements
 too much is equivalent to ignoring the LLSA completely, which would be unwise. 
We can therefore estimate the uncertainty related to the assumptions on the LLSA error 
by  varying $\xi$ between 0.7 and 1.3, obtaining the  
relative variations on the main parameters %that  we get in this way are 
\begin{eqnarray}
\delta^{\xi} V_{cb}=^{+0.2\%}_{-0.2\%}, \quad \delta^{\xi}\mu^2_{\pi}=^{+4.7\%}_{-2.0\%} , \quad \delta^{\xi}\mu^2_{G}=^{+1.0\%}_{-0.9\%}, \nonumber \\
\delta^{\xi}\rho^3_{D}=^{+18.2\%%+15.5\%
}_{-10.0\% %-11.0\%
}, \qquad \delta^{\xi}\rho^3_{LS}=^{-1.3\%}_{+0.9\%}. \qquad\
\label{eqn:deplambda}
\end{eqnarray}
We will  include this uncertainty in the final error on $\vcb$.
We also vary ${\epsilon}$ over the range $0.4 \pm 0.1 \mbox{GeV}$ to gauge the related uncertainty. The dependence of the parameters on the choice of excitation energy can be seen in Fig.~\ref{fig:epsilonplots}, and the resulting relative uncertainties are
\begin{eqnarray}
\delta^{{\epsilon}}V_{cb}=^{+0.04\%}_{-0.04\%}, \quad \delta^{{\epsilon}}\mu^2_{\pi}=^{+0.7\%}_{-0.8\%}, \quad \delta^{{\epsilon}}\mu^2_{G}=^{-0.4\%}_{+0.3\%}, \nonumber \\
\delta^{{\epsilon}}\rho^3_{D}=^{+3.3\%}_{-3.6\%}, \qquad \delta^{{\epsilon}}\rho^3_{LS}=^{+0.3\%}_{-0.4\%}, \qquad \, \,
\end{eqnarray}
which are mostly negligible. %, with the partial exception of $\rd$. 

We also repeated the default fit in two slightly different ways: $i)$ adding the PDG constraint on $m_b$  \cite{pdg} after a scheme conversion, $m_b^{kin}=4.550(42)$GeV, which leads to  $|V_{cb}| =42.10(73)\ 10^{-3}$; $ii)$ changing, in addition to that, the $m_c$ constraint into
$m_c(2{\rm GeV})=1.091(14)$GeV, obtained evolving 
the result of  \cite{Chetyrkin:2009fv} to 2GeV. This leads to a somewhat better convergence of the perturbative series for the semileptonic width \cite{Gambino:2011cq};  in this case $|V_{cb}| =42.00(64)\ 10^{-3}$ and $\chi^2/dof=0.44$. The results of all these fits are remarkably 
consistent with each other.

\section{The Fourth Hadronic Moment}
The central hadronic moments  are sensitive probes of  power corrections.
For instance, ${\cal O}(1/m_b^{4,5})$ affect $h_3$ in a significant way and one could expect even higher moments to be able to constrain the higher power contributions
in a useful way. As DELPHI has measured $h_{4,5}$ without a cut on the lepton energy
\cite{Abdallah:2005cx}, we have computed $h_4$ to explore the possibility of 
including them in the fit, despite the high correlation with lower hadronic moments. 
The result, in GeV$^8$, is
\begin{eqnarray}
\nonumber h_4 &=& 0.15_{tree} + 15.97_{pert} + 4.23_{\mupi} + 1.81_{\as\mupi }-0.16_{\mug}\\
\nonumber  &\,+& 0.74_{\as\mug} + 2.31_{\rd} -0.10_{\rls} 
  + 3.80_{m_i} -4.91_{r_i},
\end{eqnarray}
where we have evaluated the different contributions using Table II. Perturbative contributions
are largely dominant, diluting any possible ${\cal O}(1/m_b^{4,5})$ effect and amplifying the uncertainty. 
In fact, the inclusion of DELPHI's $h_4$ in the fit has negligible impact on 
$\vcb$ and the OPE parameters.

\section{Summary}
We have studied the effect of higher power corrections on the fits to inclusive semileptonic $B$ decays which determine $|V_{cb}|$. Because of the large number of new parameters at ${\cal O}(1/m_b^{4,5})$, we used the LLSA to provide loose constraints on the higher power matrix elements and performed a new global fit to the semileptonic moments.
The higher power corrections have a minor effect on $|V_{cb}|$ and on the expectation values of the lower dimensional operators, and we observe a good convergence of the heavy mass expansion.
 There is a $-0.25$\% reduction in   $|V_{cb}|$
\begin{equation}
10^3\, |V_{cb}|  = 42.11(53) (50) (07) (10) = 42.11(74), \nn
\end{equation}
where the four errors are, respectively, 
the parametric error from the fit, the theoretical error on the semileptonic width, and those due to the $\tau_B$ uncertainty, and to $\delta^\xi,\delta^\epsilon$. 
The bottom mass determination from the fit is $m_b^{kin} = 4.546(21)$\,GeV. 
A slightly more precise  alternative fit makes use of $\overline{m}_c$ at a lower scale, 2 GeV, 
and of the PDG average for $m_b$, leading to  
\begin{equation}
10^3\, |V_{cb}|  = 42.00(50) (39) (07) (10) = 42.00(64).\nn
\end{equation}
After the implementation  of various higher order effects 
the inclusive determination of $V_{cb}$ appears  robust. Further improvements 
may come from the calculation of ${\cal O}(\as/m_b^3)$  and ${\cal O}(\as^3)$ effects, from lattice QCD
determination of some of the non-perturbative parameters, and from new \cite{FBasym} and more precise 
measurements at Belle-II. 
%The ${\cal O}(\as^3)$ corrections are  challenging but technically feasible.
\begin{small}
\begin{table}[t]
\begin{tabular}{|cr||cr||cr|}
\hline
$h_2$ & -2.65 & $k_6$ & 75.20  & $k_{15}$& -34.41 \\
$h_3$ & -11.20& $k_7$ & -20.17& $k_{16}$& -17.33\\
$h_5$ & 3.12&  $k_8$ & 4.26& $k_{17}$& -0.23\\
$h_6$ & -2.94& $k_9$ & 19.91 & $k_{18}$& 18.00\\
$k_1$ & -1.25& $k_{10}$& 59.21 & $a_1 $& -1.17\\
$k_2$ & -91.12& $k_{11}$& -23.57 &$a_2 $& -4.26  \\
$k_3$ & 120.83& $k_{12}$& -26.13 &$f_{\pi}$ & 0.95\\
$k_4$ & -131.94& $k_{13}$& 26.56 &$f_{G} $& -2.10\\
$k_5$ & 20.88& $k_{14}$& 5.25 & &\\ \hline
\end{tabular}
\caption{Higher-order contributions to the semileptonic width evaluated at $ r = 0.0472$.}
\label{tab:widthfxn2}
\end{table}
\end{small}

\subsection*{Acknowledgments}
S.T.  is supported by the Advanced Grant EFT4LHC of the European Research
Council and the Cluster of Excellence {\it Precision Physics, Fundamental
Interactions and Structure of Matter} (PRISMA-EXC 1098).

\section{Appendix}
The ${\cal O}(1/m_b^4)$ corrections to the $h_i$ for $E_{cut}=1$GeV and $m_{c,b}$ from Table II are (in units GeV$^{2,4,6}$)
\bea
\delta h_1&=& 0.01{\overline m}_1+0.28{\overline m}_2+0.54{\overline m}_3-0.40
{\overline m}_4-0.04 {\overline m}_5\nn\\
&&-0.21{\overline m}_6-0.01{\overline m}_7-0.08{\overline m}_8
+0.00{\overline m}_9\nn\\
\delta h_2&=& 0.6{\overline m}_1-3.3{\overline m}_2-2.0{\overline m}_3-0.0
{\overline m}_4+0.2 {\overline m}_5\nn\\
&&+0.9{\overline m}_6+0.8{\overline m}_7+1.0{\overline m}_8
-0.2{\overline m}_9\label{dhi}\\
\delta h_3&=& -9.5{\overline m}_1+27.2{\overline m}_2-0.8{\overline m}_3+3.6
{\overline m}_4+0 {\overline m}_5\nn\\
&&+1.5{\overline m}_6-3.3{\overline m}_7-4.2{\overline m}_8
+0.6{\overline m}_9\nn. 
\eea
The total semileptonic width can be written as 
\begin{footnotesize}
\bea
\nonumber \Gamma&=& \Gamma_0 \!\left[z(r) \!\left(1\!-\!\frac{\mupi-\mug}{2m_b^2} -
\frac{\rd+\rls}{2m_b^3}+ \frac{\frac18 \overline{m}_1+\frac13 \overline{m}_4+\frac14 \overline{m}_8}{m_b^4}\right)\right.\\
\nonumber  &-&2(1-r)^4 \left( \frac{\mug}{m_b^2}-
\frac{\rd+\rls}{m_b^3} +\frac{16}{9} \frac{\overline{m}_9}{m_b^4}\right)\\
&+& d(r) \left( \frac{\rd}{m_b^3} - \frac{ 2 \overline{m}_4+\frac{2}{3} \overline{m}_9}{m_b^4} \right) +\sum_{i=2,3,5,6} h_i(r) \frac{\overline{m}_i}{m_b^4}\\
&+&\left.\sum_{i=1}^{18} k_i(r) \frac{r_i}{m_b^5}
+\sum_{i=1}^{2} a_i(r) \left(\aspi\right)^i+\sum_{i=\pi,G} f_i(r)\aspi \frac{\mu^2_i}{m_b^2} +\mbox{...} \right], \nonumber 
\eea
\end{footnotesize}
\!\!where $\Gamma_0=A_{ew}G_F^2 (m_b^{kin})^5\vcb^2/192\pi^3$, $A_{ew}=1.014$, $r = (\overline{m}_c(3{\rm GeV})/m_b^{kin})^2$ , $z(r)=1 - 8 r + 8 r^3 - r^4 - 12 r^2 \ln r$, $d(r)=2  (17 - 16 r - 12 r^2 + 16 r^3 - 5 r^4 + 12 \ln r)/3 $, and $h_i, k_i, a_i$ and $f_i$ are listed in Table~\ref{tab:widthfxn2} for a specific $r$ value. Using the values of the parameters given in Table II one gets
\begin{eqnarray}
\nonumber \frac{\Gamma}{   z(r) \Gamma_0}=  1  &-& 
0.116_{\as} -0.030_{\as^2} -   0.042_{1/m^{2}} - 0.002_{\as/m^2}
\\ &-& 0.030_{1/m^{3}} +0.005_{1/m^{4}} +0.005_{1/m^{5}}
.
\end{eqnarray}

\end{document}